\documentstyle[aps,prl,twocolumn]{revtex}
\begin{document}
\title{Quantum Clock Synchronization Based on Shared Prior
Entanglement}
\author{Richard Jozsa$^2$, Daniel S. Abrams$^1$, Jonathan P.
Dowling$^1$, and Colin P. Williams$^1$}
\address{$^1$ Jet Propulsion Laboratory, California Institute of
Technology, 4800 Oak Grove Drive, Pasadena, California 91109-8099.\\
$^2$ Dept. of Computer Science, University of Bristol, Merchant
Venturers  Building, Woodland Road, Bristol BS8 1UB, UK.}
\date{Received: \today}
\maketitle
\begin{abstract}
We demonstrate that two spatially separated parties (Alice and
Bob) can utilize shared prior quantum entanglement, and classical
communications, to establish a synchronized pair of atomic clocks.
In contrast to classical synchronization schemes, the accuracy of
our protocol is independent of Alice or Bob's knowledge of their
relative locations or of the properties of the intervening medium.

\end{abstract}
\vspace{0.5cm} In the Special Theory of Relativity, there are two
standard methods for synchronizing a pair of spatially separated
clocks, A and B, which are at rest in a common inertial frame. The
usual procedure is Einstein Synchronization (ES), which involves
an operational line-of-sight exchange of light pulses between two
observers, say Alice and Bob, who are co-located with their clocks
A and B, respectively \cite{Einstein1905}. A less commonly used
protocol  is Eddington's Slow Clock Transport. In this scheme, the
two clocks A and B are first synchronized locally, and then they
are transported adiabatically (infinitesimally slowly) to their
final separate locations \cite{Eddington1924,Anderson1998}. A
quantum algorithm for efficient clock transport has recently been
proposed by Chuang \cite{chuang}.

In this paper we propose a third protocol that utilizes the
resource of shared prior entanglement between the two
synchronizing parties. Our proposed method of Quantum Clock
Synchronization (QCS) has features in common with Ekert's
entanglement-based quantum key-distribution protocol
\cite{Ekert1991} in which Alice and Bob initially share only
prior-entangled qubit pairs. The key does not exist initially but
is created from the ensemble of entangled pairs through a series
of measurements and classical messages. Similarly for our QCS
protocol below, no actual clocks exist initially but rather only
``entangled clocks'' in a global state which does not evolve in
time. The synchronized clocks are then extracted via measurements
and classical communications performed by Alice and Bob. In this
way our QCS scheme establishes synchrony without having to
transport {\em timing} information between Alice and Bob. In
contrast, in classical synchronization schemes, actual timing
information must be transmitted from Alice to Bob over some
channel, whose imperfections generally limit the accuracy of the
synchronization.

We begin by reviewing the Ramsey temporal interferometer method
\cite{Ramsey1969} for the construction of a quantum clock. A clock
is constructed from an ensemble of two-level systems (qubits)
whose temporal evolution properties will determine the time
standard. In general any physical qubit may be used. For example,
in the International System of Units (SI),  the second is defined
as the duration of exactly 9,192,631,770 periods of oscillation
corresponding to the hyperfine (radio) transition frequency for
the ground-state of the Cs$^{133}$ atom \cite{Audoin1992}. Below
we also consider other possible physical realisations of the
qubit.

Consider a qubit with stationary states $|0\rangle$ and
$|1\rangle$ having energy eigenvalues $E_0 < E_1$ respectively. We
introduce the dual basis $|pos\rangle =
\frac{1}{\sqrt{2}}(|0\rangle + |1\rangle )$ and $|neg\rangle =
\frac{1}{\sqrt{2}}(|0\rangle - |1\rangle )$ and write $\Omega
=\frac{1}{\hbar} ( E_{1} - E_{0} ) $. The Hadamard transform $H$
is defined by the operation $|0\rangle \rightarrow |pos\rangle$
and $|1\rangle \rightarrow |neg\rangle$. Let us write $\sigma_3$
for the measurement in the $\{ |0\rangle , |1\rangle \}$ basis and
$\sigma_1$ for the measurement in the dual basis. Thus if the
qubit is a spin $\frac{1}{2}$ particle in a $z$-oriented magnetic
field then $ |0\rangle$ and $ |1\rangle $ are the $z$ spin
eigenstates with $\sigma_3$ and $\sigma_1$ being $\sigma_z$ and
$\sigma_x$ respectively. If the qubit comprises two hyperfine
energy levels of a Cs$^{133}$ atom, then $\sigma_3$ measures
population in these levels and $\sigma_1$ is measured by first
applying $H$ and then measuring $\sigma_3$.

The Ramsey method for providing a time standard is based simply on
the fact that the states $|pos\rangle$ and $|neg\rangle$ are not
stationary states. They evolve in time as:
\begin{equation} \label{evol} \begin{array}{lcl}
|pos(t)\rangle & = & \frac{1}{\sqrt{2}} \left( e^{-i \Omega t/2}
|0\rangle + e^{i \Omega t/2} |1\rangle \right)\\ |neg(t)\rangle &
= & \frac{1}{\sqrt{2}} \left( e^{-i \Omega t/2} |0\rangle - e^{i
\Omega t/2} |1\rangle \right) \end{array} \end{equation} At some
time $t=0$ we apply $H$ to an ensemble of qubits in state
$|0\rangle$ giving an ensemble of states $|pos\rangle$ which begin
to evolve as in eq. (\ref{evol}). After a time $t$ we measure the
observable $\sigma_1$ (either directly or by first applying $H$
and measuring $\sigma_3$, depending on the physical implementation
of the qubits). A straightforward calculation shows that the
probabilities for seeing outcomes 0 or 1 are given by
\begin{equation}\label{probs}
P(0) =  \frac{1}{2} \left(1 + \cos \left( \Omega t \right)
\right), \hspace{2mm} P(1) =  \frac{1}{2} \left(1 - \cos \left(
\Omega t \right) \right)
\end{equation}
By monitoring the oscillations of either $P(0)$ or $P(1)$ as a
function of time we get an estimate of the clock phase $\Omega t ~
\textrm{mod}~ 2 \pi $ and hence of $t$.

We now describe our proposed QCS scheme. We assume at the outset
that Alice and Bob share an ensemble of singlet states $|\psi^-
\rangle = \frac{1}{\sqrt{2}} \left( |0\rangle_{A} |1\rangle_{B} -
|1\rangle_{A} |0\rangle_{B}\right)$ where the subscripts refer to
particles held by Alice and Bob. The pairs are labelled $n=1,2,3,
\ldots $ and the labels are known to both Alice and Bob. This
singlet state is a ``dark state'' that does not evolve in time
provided A and B undergo identical unitary evolutions. Indeed for
{\em any} 1-qubit unitary $U$ we have $(U \otimes U) |\psi^-
\rangle = (\det U) |\psi^- \rangle$ so that $|\psi^- \rangle$
changes only by an overall unobservable phase. Our protocol below
(slightly modified) would work equally well using the state
\begin{equation} \label{delta}
|\psi^- (\eta ) \rangle = \frac{1}{\sqrt{2}} \left( |0\rangle_{A}
|1\rangle_{B} - e^{i\eta} |1\rangle_{A} |0\rangle_{B}\right)
\end{equation} for any fixed $\eta$. This state still has the
essential property of being constant in time i.e. invariant under
$U\otimes U$ where $U$ is time evolution, diagonal in the $\{
|0\rangle$,$|1\rangle \}$ basis (but unlike the singlet, it is not
invariant under $U\otimes U$ for more general $U$'s).

We will refer to a pair of clocks in the singlet state $|\psi^-
\rangle$ as an entangled pair of pre-clocks. Since
$|\psi^-\rangle$ is constant in time the pre-clock pairs could be
said to be ``idling'' -- they can provide no direct timing
information. We may also write $|\psi^-\rangle$ in the $\sigma_1$
measurement basis as
\begin{equation} \label{six}
|\psi^-\rangle = \frac{1}{\sqrt{2}}\left( |pos\rangle_A
|neg\rangle_B - |neg\rangle_A |pos\rangle_B \right) \end{equation}


Let $t$ be a time coordinate in the common rest frame of Alice and
Bob. To start the clocks at some time $t=0$, Alice simultaneously
measures all of her pre-clock pairs in the $\sigma_1$ basis $\{
|pos\rangle, |neg\rangle \} $. Thus each pair collapses randomly
and simultaneously at $A$ and $B$ into one of the following
states:
\begin{equation}\label{onetwo} \begin{array}{lcl}
|\psi^{I}\rangle &=& |pos\rangle_{A}|neg\rangle_{B} \\
|\psi^{II}\rangle &= &|neg\rangle_{A}|pos\rangle_{B} \end{array}
\end{equation}  with equal probability $\frac{1}{2}$.
The $A$ and $B$ clocks begin to evolve in time, in accordance with
Eq. (\ref{evol}) -- all starting synchronously at a time $t = 0$
in Alice and Bob's shared inertial frame. Indeed Alice's
measurement effectively reproduces the result of the first
one-clock Hadamard transform in the Ramsey scheme. However the
result here is a mixture of two equally weighted sub-ensembles I
and II. As a result of her measurement, Alice knows the labels
belonging to the subensembles I and II but Bob is unable to
distinguish them.

The density matrix of Bob's overall ensemble is still $\rho =
\frac{1}{2}I$, independent of $t$, so no measurement statistic can
provide Bob with any timing information. For Bob to extract a
clock, a classical message from Alice is required. Alice
post-selects from her entire ensemble the sub-ensemble of Type-I
qubits. Since the qubits are labelled, she can then tell Bob which
subset of his qubits are also Type-I by broadcasting their labels
via any form of classical communiqu\'e. Bob is then able to
extract his own Type-I and Type-II subensembles. Choosing the
Type-II subensemble, Bob will have a clock ensemble exactly in
phase with a Type-I clock that Alice started at $t=0$. To
establish synchrony, Bob measures $\sigma_1$ on this ensemble
(either directly or by applying $H$ and measuring $\sigma_3$) and
monitors the oscillations of $P(0)$ as in eq. (\ref{probs}). Alice
and Bob now have clocks that are ticking in unison.

The protocol as described above is still incomplete \cite{cirac}
because of the following rather subtle point: there are extra
hidden assumptions in the requirement that Alice and Bob are both
able to perform the {\em same} $H$ operation and identify the same
$|pos\rangle$ states. Indeed if we are given only $|0\rangle$ and
$|1\rangle$ as physical states (i.e. normalised vectors up to
overall phase) then the physical states $|pos\rangle$ and
$|neg\rangle$ are not uniquely determined\footnote{Note that if
$|0\rangle$ and $|1\rangle $ are given as {\em vectors} then all
other vectors such as $|pos\rangle$ {\em are} uniquely defined so
our ambiguity depends essentially on the fact that a physical
state is not just a (normalised) {\em vector} but rather, a {\em
set} of all such vectors that differ by an overall phase.} and so
$H$ is also not uniquely determined (as, for example, it entails
knowledge of $|pos\rangle$). A further arbitrary choice needs to
be made, analogous to a choice of reference frame, to fix these
further constructs.

The need for a further choice is most clearly seen by considering
the spin $\frac{1}{2}$ qubit \cite{popescu}. The physical states
$|0\rangle$ and $|1\rangle$ define a $z$ direction and
$|pos\rangle$ defines a perpendicular $x$ direction. But given
only a $z$ direction we are free to choose any orthogonal
direction as $x$. On the Bloch sphere $|0\rangle$ and $|1\rangle$
are two given poles and $|pos\rangle$ may be arbitrarily chosen to
be any point on the equator. Once $|pos\rangle$ is chosen, it must
be consistently used in all future operations. Furthermore, there
is then no further ambiguity in the identity of any state on the
Bloch sphere e.g. $|neg\rangle$ and $H$ are then uniquely fixed.

The same remarks apply to the Cs atom qubit but the physical
interpretation is quite different. The operation $H$ (and hence
$|pos\rangle$) is physically defined in terms of a $\pi/2$ pulse
applied to the physical state $|0\rangle$. But this pulse has an
origin of phase which must be chosen and then fixed (``locked")
for all future applications of $H$. Different choices of phase
locking correspond to different choices of points on the Bloch
sphere equator for $|pos\rangle$. Note that a choice of phase
locking here corresponds physically to a choice of {\em time
origin} in contrast to the spin $\frac{1}{2}$ qubit, where the
choice was a {\em spatial direction}.

For our QCS protocol to work correctly, Alice and Bob must use the
{\em same} choice of physical state $|pos\rangle$ (or equivalently
use the same choice of Hadamard operation $H$). If they use two
different choices (and use them locally consistently) then their
clocks will not be ticking in synchrony, but be offset by an
amount depending on the angle between the two choices of
$|pos\rangle$ on the Bloch equator.

In the physical implementation given by the Cs atom qubit, a
consistent choice of $H$ requires that Alice and Bob have mutually
phase locked pulses. But this is equivalent to them having clocks
ticking in synchrony thus defeating the purpose of the protocol!
However the following extension of our protocol gets around this
difficulty, allowing Alice and Bob to establish time synchrony
without the resource of mutually phase locked pulses: we duplicate
our protocol above for two different values $\Omega_1$ and
$\Omega_2$ of $\Omega$ e.g. we use two different species of atoms.
Thus Alice and Bob will require two different kinds of pulses for
the two frequencies. In his laboratory, Bob is able to lock the
phases of his two pulses, and similarly for Alice, so there will
be a common offset $\delta$ between the two locked settings of
Alice and Bob. By measuring populations in state $|0\rangle$ as
before, Alice will have the oscillations:
\begin{equation}\label{aliceosc}
P_1^A = \frac{1}{2}( 1+\cos \Omega_1 t )\hspace{5mm} P_2^A =
\frac{1}{2}( 1+\cos \Omega_2 t ) \end{equation} and Bob's
oscillations will be offset by the {\em constant} (unknown)
$\delta$:
\begin{equation} \label{bobosc}
P_1^B = \frac{1}{2}( 1+\cos (\Omega_1 t + \delta ) )\hspace{5mm}
P_2^B = \frac{1}{2}( 1+\cos (\Omega_2 t +\delta) ) \end{equation}
But now by observing the beats between the two oscillations $P_1$
and $P_2$ Alice and Bob are able to establish synchronously
ticking clocks. Indeed we have
\begin{equation} P_1^B -P_2^B = \sin ( \frac{1}{2} (\Omega_1 -
\Omega_2 )t)\sin ( \frac{1}{2} (\Omega_1 + \Omega_2 )t+\delta )
\end{equation}
so that the envelope (given by the first term) oscillates
independently of $\delta$ exactly in phase with Alice's
corresponding envelope.

It is interesting to consider the above problem, of locally
consistent but different choices of $|pos\rangle_A$ and
$|pos\rangle_B$, in the alternative physical scenario of clocks
given by ensembles of spin $\frac{1}{2}$ qubits in a magnetic
field. Although mathematically equivalent, we will see that the
physical implications are quite different. In this scenario
$|0\rangle$ and $|1\rangle$ are the $z$ spin eigenstates. We
imagine that a third party (Clare) prepares an ensemble of pairs
in the singlet state and simultaneously puts each spin in a
labelled box containing a constant magnetic field $B_z$ in the $z$
direction. She then distributes the boxes (complete with their
magnetic fields) to Alice and Bob (appropriately for each pair).
Note that Alice and Bob may determine the $z$ direction (if they
do not already know it) by measuring the direction of the
(classical) magnetic field in a box (without disturbing the
particle).

Alice now chooses an $x$ direction (perpendicular to $z$) and at
some time $t=0$ she measures $\sigma_x$ on all her particles.
Then, just as before, Bob may establish synchrony by monitoring
the oscillations of $\sigma_x$ measurement outcomes on a
sub-ensemble of his particles (selected by classical information
from Alice). The previous problem of consistent phase locked
pulses becomes the problem of Bob choosing the same $x$ direction
that Alice used. Previously, the problem was equivalent to the
original goal of the protocol (time synchrony) but here it is
different i.e. a requirement of space parallelism (``space
synchrony"). This allows the possibility of new physical
resolutions of the problem, not available for Cs atom qubits. For
example, Alice and Bob may have a prior agreement to use the
direction to the pole star as their $x$ direction (which would be
parallel to high accuracy for any two locations on Earth) i.e.
$x$-``space synchrony" may be given for free, whereas time
synchrony is not.

The idea of the previous resolution -- using two frequencies --
may be used for spin $\frac{1}{2}$ qubits as well (e.g. if Alice
and Bob are unable to see any fixed stars.) Clare sets up boxes
with two different magnetic fields (both in the $z$ direction)
giving the two different frequencies. Bob chooses his $x$ axis
randomly (perpendicular to $z$) and the constant phase offset
$\delta$ now arises from the fixed angle between Alice's and Bob's
chosen $x$ directions. An important point here is that different
physical realisations of a qubit -- although mathematically
equivalent -- lead to quite different avenues for getting around
limitations of a (mathematically) given protocol.

For some applications, such as satellite-based Very Long Baseline
Interferometry (VLBI) \cite{Levy1987}, the fact that Alice and
Bob's clocks are phase locked up to only modulo $2\pi$ is
sufficient. However, there are other applications, such as the
synchronization of satellite-borne atomic clocks in the Global
Positioning System (GPS) \cite{GPSSystem1995}, where it is
important to have a shared origin of time. For such applications,
we may adapt our QCS protocol to construct a common temporal point
of reference as follows. Using the protocol Alice and Bob set up
clocks ticking synchronously for two different frequencies
$\Omega_1$ and $\Omega_1 + \Delta \Omega$. The envelope of beats
between these frequencies oscillates with frequency
$\frac{1}{2}\Delta \Omega$. If the protocol for establishing the
two ticking synchronisations is completed in time $T$ and $\Delta
\Omega$ is chosen so that $\Delta \Omega\, T < \frac{\pi}{2}$ then
Alice and Bob may determine a unique common time origin as the
first maximum of the beat oscillations.

There are several immediate applications and advantages of our QCS
protocol. For example, in the GPS satellite constellation, the
ability of the space-borne atomic clocks to synchronize with a
master atomic clock on the ground is affected by the fluctuating
refractive index of the atmosphere, causing random variations in
the speed of light and limiting the accuracy of the classical ES
protocol. This index fluctuation error is the current limiting
factor of GPS precision \cite{GPSSystem1995}. With our QCS scheme,
the properties of the atmosphere have no effect. In fact, Alice
and Bob need not even have exact knowledge of their relative
locations.

Also classical ES requires the exchange and timing of light
pulses, but light is actually a quantum field. Hence the arrival
time of a light pulse is itself subject to quantum fluctuations,
limiting the accuracy of the ES protocol\cite{Jaekel1996}. In
contrast, our QCS scheme is unaffected by this kind of noise.

The Ramsey two-pulse temporal interferometer is isomorphic, via
the SU(2) algebra, to an optical or matter-wave Mach-Zehnder
interferometer\cite{Dowling1998}. Hence, the QCS protocol may be
readily adapted to the task of phase locking a pair of spatially
separated optical or atom interferometers, with applications to
various forms of interferometry such as VLBI.

A shortcoming of our QCS protocol that it does not specify a
method by which the shared prior entanglement between Alice and
Bob may be established. One possibility is for Alice and Bob to
meet at a common location, create an ensemble of $N$ identical EPR
pairs each in the state $\frac{1}{\sqrt{2}} \left(
|0\rangle_{A}|1\rangle_{B} - |1\rangle_{A} |0\rangle_{B}\right)$
and then go their separate ways. But then, could not Alice and Bob
just establish time synchrony at their meeting and retain accurate
clocks for future use, instead of carrying the entanglement? In
practice, clocks drift and periodic corrections of synchronisation
will be necessary, suffering from the limitations of the classical
schemes. It is not clear whether the task of carrying and
maintaining the required entanglement is equivalent to the task of
carrying and maintaining accurately running clocks. One difference
is that in the former case, the time synchrony does not initially
exist but is set up only when required, which may have
applications for security.

An alternative scheme for establishing the shared prior
entanglement would not require Alice and Bob to meet at all.
Instead, it would involve Alice and Bob each receiving
corresponding members of EPR pairs from some common source and
then using entanglement purification\cite{purify} to distill them
into an ensemble of singlet states as required by QCS.
Unfortunately, there is a hidden assumption of simultaneity in the
actions to be performed by Alice and Bob in the current
entanglement purification protocols when the states $|0\rangle$
and $|1\rangle$ are non-degenerate in energy
\cite{PreskillPrivate}, as required in our protocol. This means
ultimately that the existing entanglement purification schemes can
only create states of the form $|\psi^-(\eta )\rangle$, where
$\eta$ is unknown, rather than the true singlets (or states with
known $\eta$) needed for QCS. We are currently investigating
whether we can use such states in a modified version of QCS or
indeed whether there are alternative (asynchronous) entanglement
purification protocols that can produce pure singlets.

A second limitation of our protocol is the requirement that Alice
and Bob be relatively at rest. In a more realistic scenario we
would need to assess and correct the effects of relative motions
and accelerations, especially on the exact form of the
entanglement existing between Alice and Bob.

In conclusion, we have presented a quantum protocol for
synchronizing spatially separated atomic-clocks, which uses only
shared prior entanglement and a classical channel. The two
synchronizing parties may be at far-distant and unknown relative
locations and the accuracy of the time synchronisation is not
affected by the distance of separation or by noise on the
classical channel. Our protocol has direct applications for use in
very long baseline interferometry and also provides a means for
phase locking remote optical or matter-wave interferometers.

\begin{acknowledgements}
We would like to acknowledge valuable discussions with S. L.
Braunstein, N. Cerf, J. I. Cirac, H. J. Kimble, N. Linden, H.
Mabuchi, L. Maleki, S. Popescu, J. P. Preskill, and N. Yu.
\\D.S.A., J.P.D. and C.P.W. are supported by a contract with the
National Aeronautics and Space Administration. R. J. is supported
by the U.K. Engineering and Physical Sciences Research Council.
\end{acknowledgements}

\end{document}